# The cc-pV5Z-F12 basis set: reaching the basis set limit in explicitly correlated calculations

Kirk A. Peterson,*[a] Manoj K. Kesharwani,[b] and Jan M.L. Martin*[b]

(a) Department of Chemistry, Washington State University, Pullman, WA 99164-4630. Email: kipeters@wsu.edu FAX: +1 509 335 8867

(b) Department of Organic Chemistry, Weizmann Institute of Science, 76100 Reḥovot, Israel. Email: gershom@weizmann.ac.il. FAX: +972 8 934 4142

ABSTRACT. We have developed and benchmarked a new extended basis set for explicitly correlated calculations, namely cc-pV5Z-F12. It is offered in two variants, cc-pV5Z-F12 and cc-pV5Z-F12(rev2), the latter of which has additional basis functions on hydrogen not present in the cc-pVnZ-F12 (n=D,T,Q) sequence. A large uncontracted "reference" basis set is used for benchmarking. cc-pVnZ-F12 (n=D–5) is shown to be a convergent hierarchy. Especially the cc-pV5Z-F12(rev2) basis set can yield the valence CCSD component of total atomization energies (TAEs), *without any extrapolation*, to an accuracy normally associated with aug-cc-pV{5,6}Z extrapolations. SCF components are functionally at the basis set limit, while the MP2 limit can be approached to as little as 0.01 kcal/mol without extrapolation. The determination of (T) appears to be the most difficult of the three components and cannot presently be accomplished without extrapolation or scaling. (T) extrapolation from cc-pV{T,Q}Z-F12 basis sets, combined with CCSD-F12b/cc-pV5Z-F12 calculations appears to be an accurate combination for explicitly correlated thermochemistry. For accurate work on noncovalent interactions, basis set superposition error with the cc-pV5Z-F12 basis set is shown to be so small that counterpoise corrections can be neglected for all but the most exacting purposes.

Table of Contents Graphic (if desired):

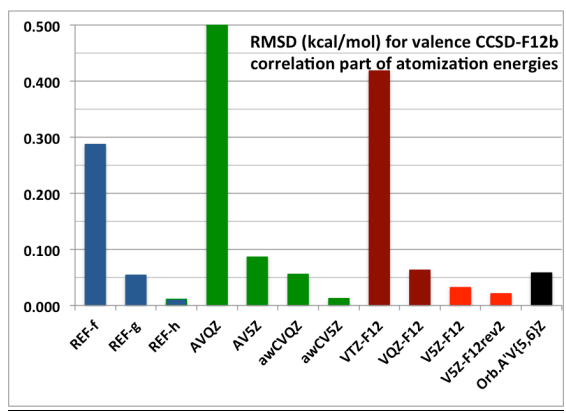

**Introduction**

Wavefunction *ab initio* calculations are for the most part a two-dimensional convergence problem (the "Pople diagram" [1]) with basis set convergence on one axis and the electron correlation method (a.k.a. n-particle treatment) on the other. For heavy elements, relativity[2] becomes the third major dimension, leading to the Császár cube.[3]

For systems with mild nondynamical correlation, the "gold standard" CCSD(T)[4,5] coupled cluster[6] method is quite close to the full configuration interaction limit, leaving basis set convergence as the main bottleneck. With conventional one-particle basis sets, convergence is quite slow,[7–9] although its comparatively slow and monotonic character naturally suggests extrapolation procedures when correlation consistent basis sets are used (e.g.,[10–12]).

Explicitly correlated methods,[13–20] in which some "geminal" terms that explicitly depend on the interelectronic distance are added to the 1-particle basis set, exhibit greatly accelerated basis set convergence. (Early work, including his own, has been reviewed by Handy;[21] for later reviews, see Refs.[22–26])

A number of forays into explicitly correlated computational thermochemistry were recently made by the present authors[27–30] and by others.[31,32] The somewhat disappointing conclusion emerged that, while F12 methods allow one to quickly reach the general neighborhood of the basis set limit, approaching more closely in a consistent way becomes quite challenging. Specifically, while basis set convergence in pure one-particle approaches (e.g., in W4 theory,[33,34] the HEAT approach,[35,36] and the FPD approach[30,37–40]) tends to be smooth with monotonically increasing TAEs (total atomization energies), basis set convergence of coupled cluster TAEs from F12 methods can be oscillatory or even (anomalously) monotonically decreasing.

One reason for this problem resides in the use of basis sets not specifically optimized for explicitly correlated calculations. A team involving one of us sought to address this problem by the development of the cc-pVnZ-F12 and cc-pCVnZ-F12 basis sets[41–43] (n=D,T,Q), which were optimized at the MP2-F12 level in the presence of the appropriate geminal terms. These basis sets do appear to have smoother convergence behavior than orbital-optimized aug-cc-pVnZ basis sets (n=D,T,Q,5,6) but obviously do not reach as far, and basis set extrapolation was still found to be necessary.[44] A crude "rule of thumb" has emerged[22–24] from practical applications, namely, that F12 calculations gain the user between two and three "zetas" over their conventional counterparts.

If that is so, expanding the cc-pVnZ-F12 by an additional member (n=5) may bring us within subchemical accuracy of the basis set limit, and indeed altogether eliminate the need for basis set extrapolation. In the present paper, we will develop and present such a basis set for H and B–Ne, and by thorough benchmarks show that it is indeed sufficiently close to the basis set limit for this purpose.

**Computational details**

Most calculations were carried out using MOLPRO 2012.1[45] running on the Faculty of Chemistry HPC cluster at the Weizmann Institute of Science. Some additional CCSD(F12) and open-shell atomic CCSD(F12*) calculations were carried out using TURBOMOLE[46] 6.4 on the same hardware.

For the cc-pVnZ-F12 correlation consistent basis sets (n=D,T,Q) optimized for F12 calculations,[41] we employed the auxiliary basis sets[47] and CABS (complementary auxiliary basis sets)[48] developed for use with them, as well as the Weigend[49,50] JK-fitting basis sets which are the MOLPRO default. Unless indicated otherwise, the geminal exponent values β recommended in Ref.[44] (rather than those of Ref.[41]) were used: β =0.9 for cc-pVDZ-F12, β =1.0 for cc-pVTZ-F12, and β =1.0 for cc-pVQZ-F12. The SCF component was improved through the "CABS correction".[19,51]

For the REF basis sets (see below), we employed very large uncontracted auxiliary basis sets previously reported in Ref.[44]. For the V5Z-F12 basis set, we considered two combinations of auxiliary basis sets: the first is that used for the REF basis sets, the second is the combination of Weigend's aug-cc-pV5Z/JKFIT basis set[50] for the Coulomb and exchange elements with Hättig's aug-cc-pwCV5Z/MP2FIT basis set[52] for both the RI-MP2 parts and for the complementary auxiliary basis set (CABS). For comparison, some F12 calculations were run with ordinary aug-cc-pVnZ basis sets,[53] where the JKFIT basis set[50] was extended by a single even-tempered layer of diffuse functions, RI-MP2 basis set was again taken from Ref.[52] but the CABS basis sets of Yousaf and Peterson[54] were employed.

Our discussion focuses on the CCSD-F12b approximation,[19,20] but we also performed some exploratory calculations using the CCSD(F12*) (a.k.a., CCSD-F12c)[55] approximation. F12 approaches as presently practiced do not directly affect the connected quasiperturbative triples, so the basis set convergence behavior of the (T) contribution is effectively that of a conventional calculation. Marchetti and Werner[56] proposed convergence acceleration by scaling the (T) contribution by the MP2-F12/MP2 correlation energy ratio, and found that this considerably improves calculated interaction energies for noncovalent complexes. Such scaling will be indicated by the notation (T*) instead of (T). (In two recent studies,[57,58] we found it to be beneficial for F12 harmonic frequency calculations and for noncovalent interaction energies[58] as well.)

The MP2-F12 correlation energies discussed are those obtained with the 3C ansatz[18] with fixed amplitudes,[16] a.k.a. "3C(Fix)".

The "frozen core" approximation was applied, i.e., all inner-shell orbitals were constrained to be doubly occupied.

For reference purposes we took the large even-tempered uncontracted *spdfgh* basis set already used in previous work by Peterson and coworkers,[59] expanded with four additional *i* functions[44] (denoted REF). In this work, truncations at a given angular momentum will be indicated as REF-f (keeping up to *f* functions inclusive), REF-g, REF-h, and REF-i. For F12 calculations, at most REF-h is used in conjunction with very large uncontracted RI, JK, and CABS basis sets;[44] for conventional calculations, up to REF-i inclusive is considered.

For comparison, conventional orbital-based SCF, CCSD, CCSD(T) results were obtained using the aug-cc-pV5Z[53] and aug-cc-pV6Z[60,61] basis sets (AV5Z and AV6Z for short), as well as with aug-cc-pCV5Z[62,63] and aug-cc-pCV6Z[64] core-valence basis sets (ACV5Z and ACV6Z for short) and the core-valence-weighted aug-cc-pwCVQZ and aug-cc-pwCV5Z basis sets.[63] In the conventional calculations, we omitted diffuse functions on hydrogen, a common practice variously denoted aug'-cc-pVnZ,[65] jul-cc-pVnZ,[66] or heavy-aug-cc-pVnZ (e.g.,[67]). No such omission was made in the F12 calculations with ordinary aug-cc-pVnZ basis sets, which were carried out purely for comparison purposes.

The principal two datasets for benchmarking were derived from the W4-11 benchmark of Karton et al.,[68] the reference geometries having been taken from the Supporting Information of that work. The "validation set" TAE71 consists of the 71 closed-shell first-row molecules among the 140 in the W4-11 set. Out of these, we selected a TAE28 "training set" of 28 molecules for which we were able to obtain CCSD-F12b total energies with the REF-h basis set, and CCSD(T) with REF-i. They are: BF, $BH_3$, BH, BN, $C_2H_2$, $C_2H_4$, $C_2$, $CF_2$, $CH_2NH$, $CH_4$, $CO_2$, CO, $F_2O$, $F_2$, $H_2CO$, $H_2O_2$, $H_2O$, $H_2$, HCN, HF, HNC, HNO, HOF, $N_2O$, $N_2$, $NH_3$, $O_3$, and $CH_2(^1A_1)$.

An additional benchmark is considered in the guise of the A24 weak interaction dataset[69] of Hobza and coworkers; reference geometries for such sets were downloaded from the BEGDB website.[70] (We note that a subset of the S22x5[71,72] weak interaction benchmark was also very recently treated by Brauer et al.[58] with the present V5Z-F12 basis sets).

**Basis set optimization**

The development of the new cc-pV5Z-F12 orbital basis sets was similar to the previous optimizations of the n=D-Q sets.[41] In the present work the HF basis sets (s functions for H and He, s and p for B-Ne) were taken from the standard contracted cc-pV6Z (H) and aug-cc-pV6Z (He, B-Ne) basis sets.[60] Correlating functions optimized for the MP2-F12/3C total energy were then added to these HF sets: (5p3d2f1g) for H, He and (5d4f3g2h) for B-Ne. In each case the exponents were constrained to follow an even-tempered sequence. For B through Ne a single $p$ function was also uncontracted from the HF aug-cc-pV6Z primitive sets (7th most diffuse). Except for He and Ne, the optimizations were carried out on the electronic ground states of the homonuclear diatomics using the geometries of Ref.[48] As in Ref.[41], a slightly stretched (1.25 x $r_e$) geometry was utilized for $H_2$. For consistency with Ref.[41], all optimizations employed a geminal exponent of 1.4 with the reference DF and RI basis sets of Ref.[44]. (This choice of the geminal exponent keeps the optimized orbital exponents somewhat more diffuse so that the F12 factor covers the short range correlation, leaving the basis set to take care of the long range.)

An alternative series of basis sets was also determined for the hydrogen atom, denoted cc-pVnZ-F12(rev2), for n=D-5. These sets were optimized as above, but with a larger number of correlating functions in analogy to B-Ne (at TZ and above), i.e., (2p) for DZ, (3p2d) for TZ, (4p3d2f) for QZ, and (5p4d3f2g) for 5Z.

The geminal exponent β was optimized at the MP2-F12 level according to same procedure as in Ref. [44], found to be β=1.2.

**Results and discussion**

(a) the SCF component of TAE

**Table 1.** RMSDs (kcal/mol) for SCF and MP2-F12 valence correlation components of total atomization energies for the TAE28 set

|  | HF+CABS | $\beta=1.4$ MP2-F12 | β opt MP2-F12 |
|---|---|---|---|
| VTZ-F12 | 0.052 | 0.157 | 0.166 |
| VQZ-F12 | 0.012 | 0.031 | 0.050 |
| V{T,Q}Z-F12 |  | 0.038 | 0.020 |
| V5Z-F12 | 0.001 | 0.009 | 0.014 |
| V5Z-F12(rev2) | 0.001 | 0.011 | 0.011 |
| REF-f | 0.006 | 0.100 |  |
| REF-g | 0.000 | 0.019 |  |
| REF-h | 0.000 | 0.004 |  |
| awCVQZ | 0.061 | 0.050 |  |
| awCV5Z | 0.003 | 0.012 |  |
| AVQZ | 0.038 | 0.120 |  |
| AV5Z | 0.013 | 0.061 |  |
|  | Orbital-HF |  |  |
| AV5Z | 0.029 |  |  |
| AV6Z | 0.010 |  |  |
| ACV5Z | 0.008 |  |  |
| ACV6Z | 0.002 |  |  |

In orbital-based *ab initio* calculations, convergence of the correlation energy is so much slower compared to that of the Hartree-Fock energy that with the sizes of basis sets (like aug-cc-pV6Z) routinely used in such calculations, the Hartree-Fock component is typically well converged. (See however Ref.[73] for a caveat.) In explicitly correlated calculations, on the other hand, the Hartree-Fock component typically needs to be refined with a CABS (complimentary auxiliary basis set) type correction[19,51] in order to obtain the SCF component at a similar accuracy as the MP2-F12 or CCSD-F12b correlation energies.

We thus consider the total atomization energies of the TAE28 set. Our reference data are orbital SCF/REF-i calculations. RMSDs at various levels of theory are given in Table 1. As the SCF/REF-h results differ by less than 0.001 kcal/mol RMSD, we can safely consider the SCF/REF-i values to represent the basis set limit.

For conventional AV{5,6}Z, the RMSDs are 0.029, and 0.010 kcal/mol, respectively. Note, however, that switching to the core-valence versions of these basis sets, ACV{5,6}Z, drastically reduces these errors to 0.008 and 0.002 kcal/mol, respectively, illustrating the importance of additional radial flexibility for accurate SCF energies. It has been noted by both groups involved that W4 (which takes valence extrapolation as the starting point) and HEAT (which takes all-electron calculations as the starting point) sometimes yield nontrivially different values for the same molecule at the CCSD(T) limit: it appears this must be ascribed to the additional flexibility of the core-valence basis set for the SCF and valence correlation components, rather than the inner-shell correlation energy. (By way of illustration: the SCF/ACVQZ – SCF/AVQZ difference can become as large as 0.19 kcal/mol for $CO_2$ and 0.18 kcal/mol for $N_2O$.)

For HF+CABS/cc-pV5Z-F12, the RMSD is just 0.001 kcal/mol, functionally equivalent to the HF limit. For cc-pVQZ-F12, this increases to 0.012 kcal/mol, for cc-pVTZ-F12 to 0.052 kcal/mol. We can thus conclude that HF+CABS is effectively at the basis set limit for cc-pV5Z-F12 (and *a fortiori* for V5Z-F12(rev2)).

(b) the valence MP2-F12 component

RMSDs for the MP2 components for the TAE28 set can again be found in Table 1. As the reference values we carried out two-point extrapolations from REF-{g,h} values assuming an $L^{-7}$ dependence. The extrapolation covers just 0.004 kcal/mol, and even the REF-g results are within 0.02 kcal/mol RMS of the reference.

The cc-pV5Z-F12 basis set is within 0.01 kcal/mol of the reference values, on average. For comparison, the similar-sized AV5Z basis set still has an RMSD error of 0.06 kcal/mol, which drops down to 0.01 kcal/mol when core-valence functions are added. Progressively deleting those reveals that almost all the improvement resides in greater radial flexibility in the s and p spaces: d-type core-valence functions have little effect, and f- and higher basically none.

From V{T,Q}Z-F12 results and using the extrapolation of Hill et al.,[44] an RMSD of just 0.02 kcal/mol can be obtained at much lower expense: the attraction of V5Z-F12 then merely rests in being able to forgo extrapolation entirely.

(c) the valence CCSD-F12b component

Error statistics are given in Table 2; the final column, the number of basis functions for one molecule ($C_2H_4$), is given as an approximate indicator of computational cost.

**Table 2.** RMSDs (kcal/mol) for CCSD-F12b valence correlation components of total atomization energies

| | TAE28 | TAE71 | TAE71 | $C_2H_4$ |
|---|---|---|---|---|
| | — With respect to different reference levels: — | | | |
| | REF{g,h} $\beta$=1.4 | awCV5Z $\beta$=1.4 | AV{5,6}Z Schwenke | virtuals(d) |
| VDZ-F12 $\beta$=0.9 | 1.991 | 3.184 | 3.159 | 88 |
| VTZ-F12 $\beta$=1.0 | 0.589 | 0.927 | 0.900 | 170 |
| VTZ-F12 $\beta$=1.4 | 0.418 | 0.664 | 2.242 | 170 |
| VQZ-F12 $\beta$=1.0 | 0.143 | 0.223 | 0.214 | 302 |
| VQZ-F12 $\beta$=1.4 | 0.063 | 0.087 | 0.107 | 302 |
| V5Z-F12 $\beta$=1.2 | 0.052 | 0.073 | 0.090 | 496 |
| V5Z-F12 $\beta$=1.4 | 0.032 | 0.036 | 0.078 | 496 |
| V5Z-F12(rev2) $\beta$=1.2 | 0.033 | 0.044 | 0.066 | 580 |
| V5Z-F12(rev2) $\beta$=1.4 | 0.021 | 0.017 | 0.069 | 580 |
| V{T,Q}Z-F12 $\beta$=1.0 (a) | 0.125 | 0.183 | 0.234 | 302 |
| V{T,Q}Z-F12 $\beta$=1.4 (a) | 0.066 | 0.081 | 0.141 | 302 |
| Schwenke A'V{5,6}Z (c) | 0.058 | 0.071 | REFERENCE | 734 |
| awCVQZ $\beta$=1.4 | 0.056 | 0.066 | 0.096 | 394 |
| awCV5Z $\beta$=1.4 | 0.013 | REFERENCE | 0.018 | 674 |
| AVQZ $\beta$=1.4 | 0.543 | 0.443 | 0.487 | 336 |
| AV5Z $\beta$=1.4 | 0.086 | 0.139 | 0.192 | 466 |
| AV{Q,5}Z $\beta$=1.4 (a) | 0.228 | 0.154 | 0.170 | 466 |
| REF-f $\beta$=1.2 | 0.313 | | | 670 |
| REF-f $\beta$=1.4 | 0.287 | | | 670 |
| REF-g $\beta$=1.2 | 0.067 | | | 958 |
| REF-g $\beta$=1.4 | 0.054 | | | 958 |
| REF-h $\beta$=1.2 | 0.017 | | | 1244 |
| REF-h $\beta$=1.4 | 0.011 | | | 1244 |
| REF-i | — | | | 1348 |
| REF-{g,h} $\beta$=1.2 (b) | 0.004 | | | F12b (b) |
| REF-{g,h} $\beta$=1.4 (b) | REFERENCE | | | F12b (b) |

ROHF-UCCSD(T) used for atomic energies throughout.
  (a) Extrapolated by $E_\infty = E(L) + [E(L)-E(L-1)]/((L/L-1)^\alpha - 1)$ with $\alpha$ for that basis set pair taken from Ref. [44]
  (b) Extrapolated by $E_\infty = E(L) + [E(L)-E(l-1)]/((L/L-1)^7 - 1)$
  (c) conventional orbital-based CCSD calculation
  (d) indicative of basis set size. Cost can be assumed to scale as $N^4$.

As was seen repeatedly before, the CCSD valence correlation component actually displays *slower* basis set convergence in F12 calculations than the MP2 component.

For the TAE28 "training set", we were able to obtain REF-g and REF-h values with both $\beta=1.2$ and $\beta=1.4$. The RMSD between the two sets is just 0.004 kcal/mol, confirming that sensitivity toward the geminal exponent $\beta$ is very weak as long as the orbital basis is sufficiently large. We have (somewhat arbitrarily) used the $\beta=1.4$ values as our reference.

Performance of conventional AVnZ basis sets clearly is unacceptably slow through AVQZ inclusive: the RMSD for AVQZ still exceeds 0.5 kcal/mol, even as that for AV5Z drops down to 0.09 kcal/mol. In contrast, awCVQZ and awCV5Z have RMSD errors of just 0.06 and 0.02 kcal/mol: for comparison, Schwenke extrapolation[74] from conventional A'V{5,6}Z results yields RMSD=0.06 kcal/mol. Like for the HF components, stripping core-valence basis functions one angular momentum at a time revealed that the main improvement from AVQZ to awCVQZ results from greater radial flexibility in the (s,p) sets, followed by the d functions.

Considering the VnZ-F12 sequence with recommended geminal exponents {0.9,1.0,1.0,1.2}, we find RMSDs tapering off rapidly from 2.0 (D) via 0.6 (T) and 0.14 (Q) to 0.05 (n=5) kcal/mol. When instead $\beta=1.4$ (often favored in large basis set studies) was employed, these RMSDs are, at least relatively speaking, significantly reduced, to 0.4 kcal/mol for VTZ-F12 via 0.11 kcal/mol for VQZ-F12 to 0.03 kcal/mol for V5Z-F12.

By way of additional perspective, with $\beta=1.4$, REF-f reaches 0.29 kcal/mol and REF-g 0.05 kcal/mol, suggesting that VTZ-F12 and VQZ-F12 still are some distance removed from *spdf* and *spdfg* radial saturation, respectively.

The VnZ-F12(rev2) basis sets, with their additional polarization functions for hydrogen, actually yields slightly worse error statistics for VDZ-F12 (due to "error de-compensation") and only marginal improvements compared to VTZ-F12 and VQZ-F12, but for V5Z-F12(rev2) the improvement is more notable in relative terms. Especially for such species as $C_2H_4$ the difference is significant.

As REF-h is not a practical option for most species in the TAE71 set, a secondary standard needs to be chosen. In view of the very low RMSD of 0.013 kcal/mol for awCV5Z($\beta$=1.4), this will be our first choice: a second choice, less as an actual secondary standard than as a comparison point to conventional orbital-based calculation, is AV{5,6}Z with Schwenke's extrapolation.[74]

Compared to awCV5Z($\beta$=1.4), for TAE71, VnZ-F12 RMSD are 3.2, 0.93, 0.22, and 0.07 kcal/mol for n = D, T, Q, and 5, respectively; for V5Z-F12(rev2), a further lowering to 0.04 kcal/mol is seen which (naturally) derives from the many species with several hydrogens that are part of TAE71 but not its subset TAE28. For $\beta$=1.4, we can reduce these values to 0.66 for VTZ-F12, 0.09 for VQZ-F12, and 0.04 kcal/mol for V5Z-F12, with V5Z-F12(rev2) reaching a low of 0.017 kcal/mol that is similar to the uncertainty in the secondary standard itself.

What about extrapolation from V{T,Q}Z-F12 results? With the extrapolation exponent 4.596 given in Ref.[44], RMSD=0.18 kcal/mol for $\beta$=1.0 and 0.08 kcal/mol for $\beta$=1.4, which are relatively small improvements over the raw VQZ-F12 results. For V5Z-F12, extrapolation appears to be entirely redundant.

The bottom line from this entire section is that with the V5Z-F12 and especially the V5Z-F12(rev2) basis set, the valence CCSD atomization energies of small molecules can be obtained *without any extrapolation* at an accuracy comparable or superior to the conventional extrapolation techniques involving 5Z and 6Z basis sets used in methods like W4 theory,[33,34] FPD,[30,37–40] and HEAT.[35,36]

In terms of relative computational cost, V5Z-F12 or even V5Z-F12(ref2) will be an order of magnitude cheaper than REF-h. V5Z-F12(ref2) is of course equivalent to V5Z-F12 if no hydrogen atoms are present, but for e.g., ethylene, the computational cost will approximately double, but the additional energy lowering is significant in high-accuracy computational thermochemistry. It is still only about half as expensive as the awCV5Z basis set being used as the secondary standard.

As an aside, we considered the difference between CCSD-F12b[19,20] and the more rigorous CCSD(F12*)[55] (a.k.a. CCSD-F12c) method, as it converges along the cc-pVnZ-F12 series. (As MOLPRO has no open-shell CCSD-F12c implementation, the F12c-F12b differences for the atoms were calculated using TURBOMOLE.[46]) Compared to the orbital-based A'V{5,6}Z extrapolated contributions to the TAE71 atomization energies (which we estimate to be reliable to only about 0.1 kcal/mol, but obviously are not biased toward one F12 approximation or another), CCSD(F12*) is noticeably closer than CCSD-F12b for n=D (by 1.2 kcal/mol) and n=T (by 0.4 kcal/mol), but for n=Q and n=5 the RMSD values are comparable for both approaches. For molecules dominated by dynamical correlation, the differences taper off to nearly nothing for n=5. For ozone and other molecules with significant post-CCSD(T) correlation terms,[34,68] however, nontrivial differences (0.1–0.3 kcal/mol) between CCSD-F12b and CCSD(F12*) appear to remain at the basis set limit. We speculate that this may reflect the limitations of an MP2-F12 based *ansatz* for systems where MP2 is a poor approximation to the correlation energy. CCSD(F12*) entails fewer approximations than CCSD-F12b and thus should in principle be more rigorous. In practice, however, the presently available data are inconclusive.

At the request of a reviewer, we considered the effect of not neglecting the CABS contributions to the projector in the CCSD-F12 coupling term in the CCSD-F12b calculations (IXPROJ=1). For cc-pVDZ-F12, the effect on the TAE71 atomization energies is 0.48 kcal/mol RMSD, which drops to 0.048 kcal/mol for cc-pVTZ-F12 and to just 0.004 kcal/mol for cc-pVQZ-F12.

(d) the valence (T) component

**Table 3.** MSD and RMSD for the (T) components of the TAE28 atomization energies for different basis sets and approximations (kcal/mol), as well as global (T) scale factors (dimensionless) for the (Ts) approximation.

|  | MSD orbital | MSD (T)F12b | MSD (T*)F12b | MSD (Ts)F12b | RMSD orbital | RMSD (T)F12b | RMSD (T*)F12b | RMSD (Ts)F12b | Sc.fac. (Ts)F12b |
|---|---|---|---|---|---|---|---|---|---|
| REF-f | -0.171 | -0.228 | +0.179 |  | 0.199 | 0.267 | 0.237 |  |  |
| REF-g | -0.084 | -0.112 | +0.081 |  | 0.086 | 0.118 | 0.118 |  |  |
| REF-h | -0.048 | -0.063 | +0.046 |  | 0.040 | 0.058 | 0.074 |  |  |
| REF-i | -0.018 | – | – |  | 0.022 | – | – |  |  |
| REF-{h,i} | REF | – | – |  | REF | – | – |  |  |
| VDZ-F12 | – | -1.217 | +0.149 | -0.135 | – | 1.358 | 0.355 | 0.436 | 1.1413 |
| VTZ-F12 | – | -0.492 | +0.135 | -0.050 | – | 0.532 | 0.202 | 0.133 | 1.0527 |
| VQZ-F12 | – | -0.227 | +0.090 | -0.026 | – | 0.235 | 0.131 | 0.054 | 1.0232 |
| V5Z-F12 | – | -0.137 | +0.065 | -0.018 | – | 0.135 | 0.094 | 0.027 | 1.0136 |
| V5Z-F12r2 | – | -0.128 | +0.062 | -0.013 | – | 0.129 | 0.092 | 0.024 | 1.0131 |
| {T,Q} | – | -0.011 | – |  | – | 0.024 | – |  |  |
| {Q,5} | – | -0.024 | – |  | – | 0.036 | – |  |  |
| AV{5,6}Z Petersson | -0.007 | – | – |  | 0.010 | – | – |  |  |

In Table 3 can be found error statistics for the (T) component, both RMSDs (root-mean-square deviations) and (for the sake of detecting systematic bias) MSD (mean signed deviations). As this contribution does not benefit from the geminal terms, basis set convergence of the unscaled quasiperturbative triples is expected to be similar to that in a conventional calculation.

As the reference, we employ REF-h and REF-i conventional calculations, extrapolated to the infinite-basis limit using $1/L^{-3}$. The extrapolation in this case covers just 0.022 kcal/mol RMS.

It should be noted that conventional AV5Z and AV6Z calculations extrapolated using the expression of Ranasinghe and Petersson[75] agree with our reference values to 0.01 kcal/mol RMSD (–0.007 signed).

From the MSDs, it is clear that unscaled (T)-F12b corrections with the VnZ-F12 basis sets yield unacceptable systematic underestimates of (T), even with the V5Z-F12 and V5Z-F12(rev2) basis sets. Marchetti-Werner scaling,[56] indicated by the symbol (T*), leads instead to mild systematic overestimates, with the (T*) coming closer to the basis set limit as the basis set is increased. It appears to matter little for the statistics whether the atomic (T) contributions are scaled by their own $E_{corr}$[MP2-F12]/$E_{corr}$[MP2] ratios or that of the molecule, even though only the latter choice is strictly size-consistent.

Another approach would be to use a single global scale factor for each basis set, fitted to minimize the RMSD. The error statistics obtained in this manner are much more favorable, especially for the larger basis sets: for V5Z-F12 the RMSD drops to 0.027 kcal/mol and is reduced slightly further to 0.024 kcal/mol for V5Z-F12(rev2). The fitted scale factors are noticeably smaller than those obtained from $E_{corr}$[MP2-F12]/$E_{corr}$[MP2] according to the Marchetti-Werner prescription, but fairly similar to those obtained from $E_{corr}$[CCSD-F12b]/$E_{corr}$[CCSD].

What about basis set extrapolation from smaller basis sets? For V{T,Q}Z-F12 using the extrapolation exponent 2.895 given in Table XI of Ref.[44] a pleasingly low RMSD=0.024 kcal/mol for the (T) contribution is obtained. Fitting an extrapolation exponent 2.61 for V{Q,5}Z-F12, we obtain an in fact slightly poorer RMSD=0.036 kcal/mol.

All this hints at a computational thermochemistry approach in which CCSD(T)-F12b/VTZ-F12 and VQZ-F12 calculations are combined with CCSD-F12b/V5Z-F12 or, better still, CCSD-F12b/V5Z-F12(rev2).

(d) <u>Weak molecular interactions</u>

Results have already been reported for dissociation curves of a seven-system subset of the S22x5 dataset[71,72] in a previous paper[58] on the desirability of counterpoise corrections in explicitly correlated studies of weak molecular interaction. In short, we found that with the V5Z-F12 basis set we could effectively obtain basis set limit values.

**Table 4.** RMSD (kcal/mol) at the CCSD(T)-F12b level for the A24 set of weak interactions using different counterpoise choices.

|  | VDZ-F12 | VTZ-F12 | VQZ-F12 | V5Z-F12 | V5Z-F12(rev2) |
|---|---|---|---|---|---|
| With (T*) scaling | | | | | |
| raw | 0.050 | 0.036 | 0.018 | 0.012 | 0.011 |
| counterpoise | 0.113 | 0.036 | 0.007 | 0.007 | |
| half-half | 0.063 | 0.016 | 0.009 | 0.009 | |
| Without (T) scaling | | | | | |
| raw | 0.061 | 0.022 | 0.011 | 0.009 | 0.008 |
| counterpoise | 0.159 | 0.058 | 0.015 | 0.004 | |
| half-half | 0.104 | 0.027 | 0.007 | 0.005 | |

Reference data were taken from Ref.[69]

Presently, we shall consider the more recent A24 dataset of Řezáč and Hobza.[69] Two systems containing argon needed to be deleted. RMSDs with and without counterpoise corrections are given in Table 4. The reference geometries from the Ref. [69] were downloaded from begdb.com[70] and used without further optimization.

If we assume the reference values from Ref.[69] to be accurate to about 0.01 kcal/mol, then VQZ-F12 can reach that accuracy level with a counterpoise correction, while V5Z-F12 can do so even without counterpoise. For weak interaction work, at least, V5Z-F12(rev2) does not appear to offer any advantages over V5Z-F12. In fact, the values for the A24 dataset agree to ±0.001-3 kcal/mol RMS between V5Z-F12 and V5Z-F12(rev2).

**Table 5.** Comparison of RMS counterpoise corrections (kcal/mol) at the CCSD(T)/aug'-cc-pVnZ and CCSD(T)-F12b/cc-pVnZ-F12 levels for the A24 set of weak interactions.

|  | CCSD(T) |  | CCSD(T*)-F12b | CCSD(T)-F12b |
|---|---|---|---|---|
|  |  | cc-pVDZ-F12 | 0.122 | 0.120 |
| aug'-cc-pVTZ | 0.254 | cc-pVTZ-F12 | 0.065 | 0.066 |
| aug'-cc-pVQZ | 0.095 | cc-pVQZ-F12 | 0.020 | 0.023 |
| aug'-cc-pV5Z | 0.035 | cc-pV5Z-F12 | 0.006 | 0.009 |

What about basis set superposition error relative to conventional calculations? RMS values including counterpoise corrections for the A24 set with similar-sized basis sets for orbital and F12 calculations are compared in Table 5. For CCSD(T*)-F12b/cc-pVTZ-F12, the RMS BSSE is about one-quarter that for CCSD(T)/aug-cc-pVTZ; for cc-pV5Z-F12, the corresponding ratio goes down to about one-sixth, favoring the F12 calculation even further. In fact, the value of 0.006 kcal/mol for cc-pV5Z-F12 is small enough that, for most purposes, the counterpoise correction can be neglected in a CCSD(T*)-F12b/cc-pV5Z-F12 calculation.

As a further illustration, we have calculated the counterpoise corrections (CP) for $(H_2O)_2$ and $(HF)_2$ (systems 2 and 4 of the A24 set) at the CCSD(T)/aug-cc-pV6Z level. The respective CPs are still 0.028 and 0.033 kcal/mol, compared to 0.055 and 0.065 kcal/mol at the CCSD(T)/aug-cc-pV5Z level but just 0.010 and 0.011 kcal/mol at the CCSD(T*)-F12b/cc-pV5Z-F12 level.

**Conclusions**

We have developed and benchmarked a new extended basis set for explicitly correlated calculations involving H, He, and B–Ne, namely cc-pV5Z-F12. It is offered in two variants, cc-pV5Z-F12 and cc-pV5Z-F12(rev2), the latter of which has additional basis functions on hydrogen not offered in the cc-pVnZ-F12 (n=D,T,Q) sequence. These rev2 sets on H are also developed for $n$ = D-Q. A large uncontracted "reference" basis set is used for benchmarking cc-pVnZ-F12 (n=D–5) is shown to be a convergent hierarchy. Especially the cc-pV5Z-F12(rev2) basis set can yield the valence CCSD component of TAEs, *without any extrapolation*, to an accuracy normally associated with aug-cc-pV{5,6}Z extrapolations. SCF components are effectively at the basis set limit, while MP2 is within as little as 0.01 kcal/mol. However, the determination of (T) cannot presently be accomplished without extrapolation or scaling.

A composite scheme in which SCF and valence CCSD components are evaluated with the cc-pV5Z-F12(rev2) basis set and (T) extrapolated from CCSD(T)-F12b/cc-pV{T,Q}Z-F12 calculations appears to offer great promise for accurate thermochemical calculations.

In addition, the aug-cc-pwCV5Z basis set appears to be large enough that it can be used as a reference standard when CCSD-F12b/REF-h calculations are not technically feasible and still greater accuracy than CCSD-F12b/V5Z-F12(rev2) is required.

For weak interactions, cc-pV5Z-F12 effectively may be considered as within 0.01-0.02 kcal/mol of the basis set limit.

**Author Contributions**

The manuscript was written through contributions of all authors. All authors have given approval to the final version of the manuscript.

**ACKNOWLEDGMENTS**. JMLM would like to thank Drs. Amir Karton (U. of Western Australia, Perth), Sebastian Kozuch (U. of North Texas, Denton) and Martin Suhm (U. of Göttingen, Germany) for helpful discussions.

This research was supported in part by the Lise Meitner-Minerva Center for Computational Quantum Chemistry and by the Helen and Martin Kimmel Center for Molecular Design.

**Supporting Information**. V5Z-F12 and V5Z-F12(rev2) basis sets for H, B–Ne in machine-readable format.

# The cc-pV5Z-F12 basis set: reaching the basis set limit in explicitly correlated calculations

*Kirk A. Peterson,*[*a] *Manoj K. Kesharwani,*[b] *and Jan M.L. Martin*[*b]

(a) Department of Chemistry, Washington State University, Pullman, WA 99164-4630. Email: kipeters@wsu.edu FAX: +1 509 335 8867

(b) Department of Organic Chemistry, Weizmann Institute of Science, 76100 Reḥovot, Israel. Email: gershom@weizmann.ac.il. FAX: +972 8 934 4142

**Supporting Information**

### (1) cc-pV5Z-F12 basis sets

Note: [6s] contractions taken from standard cc-pV6Z basis set
  p,H,5.0009,2.1439,0.9191,0.3940,0.1689
  d,H,1.3104,0.5777,0.2547
  f,H,0.6427,0.3330
  g,H,0.5575

Note: [7s] contractions taken from standard aug-cc-pV6Z basis set
  p,He,14.1317,5.4828,2.1272,0.8253,0.3202
  d,He,4.3691,1.5154,0.5256
  f,He,0.8373,0.4064
  g,He,0.6712

Note: for B-Ne, [8s7p] contractions taken from standard aug-cc-pV6Z basis set plus uncontract the 7th most diffuse p-type function

d,B,4.3748,1.8133,0.7516,0.3115,0.1291
f,B,1.4530,0.7020,0.3391,0.1638
g,B,0.7275,0.3806,0.1991
h,B,0.4849,0.2407

d,C,7.4466,2.9061,1.1341,0.4426,0.1727
f,C,2.7869,1.2123,0.5274,0.2294
g,C,1.2535,0.5824,0.2706
h,C,0.6203,0.3007

d,N,10.6384,4.1361,1.6081,0.6252,0.2431
f,N,4.5264,1.8949,0.7933,0.3321
g,N,2.1107,0.8976,0.3817
h,N,0.9543,0.4524

d,O,11.6823,4.5148,1.7448,0.6743,0.2606
f,O,6.2191,2.5090,1.0122,0.4084
g,O,3.5330,1.3658,0.5280
h,O,2.1942,0.6648

d,F,15.9912,6.1203,2.3424,0.8965,0.3431
f,F,8.3547,3.2426,1.2585,0.4885
g,F,5.4453,1.8851,0.6526
h,F,3.7955,0.7816

d,Ne,20.2445,7.6340,2.8787,1.0855,0.4093
f,Ne,10.9600,4.1223,1.5505,0.5832
g,Ne,7.2314,2.4381,0.8220
h,Ne,5.2653,0.9830

## (2) cc-pVnZ-F12rev2 basis sets for H

Note: s-type contractions taken from cc-pVnZ-F12 throughout

!cc-pVDZ-F12rev2 (2p)
p,H,0.5983,0.2232

! cc-pVTZ-F12rev2 (3p2d)
p,H,0.9312,0.3983,0.1704
d,H,0.6127,0.2653

! cc-pVQZ-F12rev2 (4p3d2f)
p,H,1.8317,0.8025,0.3516,0.1540
d,H,0.8859,0.4521,0.2307
f,H,0.6114,0.3135

! cc-pV5Z-F12rev2 (5p4d3f2g)
p,H,2.0171,1.0383,0.5344,0.2751,0.1416
d,H,2.2760,1.0833,0.5157,0.2455
f,H,1.1854,0.6176,0.3218
g,H,1.8204,0.5131